\documentclass{PoS}
\usepackage{amsmath}
\usepackage{slashed}
\usepackage{graphicx}
\usepackage{subcaption}
\usepackage{simplewick}
\usepackage[utf8]{inputenc}
\newcommand{\arxiv}[2]{[arXiv:\,\href{http://arxiv.org/abs/#1}{\texttt{#1}} [\texttt{#2}]]}

\title{Study of thermal SU(3) supersymmetric Yang-Mills theory and near-conformal theories from the gradient flow}

\ShortTitle{SU(3) SYM and near-conformal theories from the gradient flow}

\author{Georg Bergner\\
        University of Jena, Institute for Theoretical Physics\\ Max-Wien-Platz 1, D-07743 Jena, Germany \\
        E-mail: \email{georg.bergner@uni-jena.de}}
\author{\speaker{Camilo Lopez} \\
University of Jena, Institute for Theoretical Physics\\ Max-Wien-Platz 1, D-07743 Jena, Germany \\
        E-mail:\email{camilo.lopez@uni-jena.de}}
\author{Stefano Piemonte\\
        University of Regensburg, Institute for Theoretical Physics \\ Universit\"atstr. 31, D-93040 Regensburg, Germany\\
        E-mail: \email{stefano.piemonte@ur.de}}

\abstract{We compute the renormalization group flow of the mass anomalous dimension in adjoint QCD with $N_{f}=1$, $3/2$, and 2 Dirac fermions, using the gradient flow. Preliminary results are in agreement with at least a near-conformal scenario in all cases. At the largest flavor numbers we obtain the strongest indication for an IR conformal fixed point scenario. Moreover, we provide results for the thermal phase transitions in SU(3) supersymmetric Yang-Mills theory. We find hints for a connection between chiral and center symmetries in terms of a single first order phase transition where chiral symmetry is restored and center symmetry gets broken. }

\FullConference{37th International Symposium on Lattice Field Theory - Lattice2019\\
		16-22 June 2019\\
		Wuhan, China}

\begin{document}

\section{Introduction}

The investigation of the IR phase of four dimensional asymptotically free quantum field theories is a challenging non-perturbative problem. In recent years, progress has been made in the development of analytical approaches like 't Hooft anomaly matching \cite{tHooft:1979rat,Gaiotto:2017yup}, which provide some constraints and conjectures for possible IR scenarios. Nevertheless, there is currently no method to determine which IR scenario is actually realized in a given strongly interacting gauge theory. Lattice Monte Carlo simulations are therefore required to determine, for example, whether an IR conformal scenario or chiral symmetry breaking is realized.

The knowledge of the IR phase of a quantum field theory is not only important from the purely theoretical point of view, but it is also relevant for constructions of strongly interacting extensions of the Standard Model. In this proceeding we focus on the low energy behavior of Yang-Mills theory coupled to $N_f$ fermions in the adjoint representation (QCD(adj)). The theory flows towards an interacting conformal fixed point, also known as  Banks-Zaks (BZ) fixed point, if it is inside the so-called conformal window $N_{f}^{\mathrm{l}}\leq N_{f}\leq N_{f}^{\mathrm{u}}$ with the perturbative estimate $N_{f}^{\mathrm{u}}= \frac{11}{4}$ \cite{Dietrich:2006cm}. For larger values of $N_{f}$ the theory enters in a Coulomb-like phase and asymptotic freedom is lost. IR conformality is lost for smaller number of flavors ($N_f<N_{f}^{\mathrm{l}}$) when the theory becomes QCD-like, i.e. a fermion condensate forms and the spectrum is gapped. Outside the conformal window but closer towards its lower edge, conformality is weakly broken. It is expected that the running of the coupling in a certain range is small, signaling a walking or near-conformal behavior. QCD(adj) with one Majorana fermion ($N_f=1/2$ Dirac flavors), corresponds to supersymmetric Yang-Mills theory, for which is predicted to be outside the conformal window with a QCD-like behaviour. According to lattice investigations, $N_{f}^{\mathrm{l}}$ is expected to be in the range of 1 and 2 \cite{DelDebbio:2010hx,Athenodorou:2014eua,Bergner:2016hip,Bergner:2017gzw}. The precise determination of $N_f^{\mathrm{l}}$ is rather challenging. Indeed, the finite volume and UV cutoff imposed by the lattice, together with the non-zero fermion mass implied by some discretizations, break conformal symmetry explicitly. Recently, a new approach has been proposed to get further insights concerning the IR scenario and the properties of the theory close to an IR fixed point \cite{Carosso:2018bmz}. It allows to study the renormalization group (RG) flow of scaling dimensions on the lattice, using the properties of the gradient flow (GF). This opens up the possibility to compute conformal data on the lattice numerically.

In this proceeding we present preliminary results of two ongoing projects. First we show the mass anomalous dimension $\gamma_{m}$ of SU(2) QCD(adj) with  $N_{f}=1\,,3/2$ and 2 Dirac flavours determined with the GF approach. We look for the freezing of $\gamma_{m}$ along the RG flow and we are able to recognize hints for an IR conformal fixed point at $N_{f}=3/2$ and 2. For $N_{f}=1$ we find at least near-conformal behaviour.

In the second part of this proceeding we focus on $N_{f}=1/2$, which is already known to be below the conformal window. We study the thermal phase transitions for SU(3) supersymmetric Yang-Mills theory. We compute the gluino condensate and the Polyakov loop at different temperatures. Our results are consistent with the existence of a unique first order phase transition, where, within the current uncertainties, at the same temperature the non-anomalous chiral symmetry is restored and center symmetry is broken. This suggests that chiral symmetry and quark confinement are correlated phenomena in SU(3) SYM. This observation extends our results for the SU(2) case \cite{Bergner:2019dim}. 

\section{The mass anomalous dimension of adjoint QCD}

We have determined the RG flow of $\gamma_{m}$ for SU(2) QCD(adj) using the GF. In the following we will just briefly summarize the main idea of the method \cite{Carosso:2018bmz} adapted to our investigations with adjoint Wilson fermions. We won't discuss the details about GF and its renormalization properties but refer to the original papers  \cite{Luscher:2011bx,Luscher:2013cpa}.

\begin{figure}[h!]
  \centering
  \includegraphics[scale=0.6]{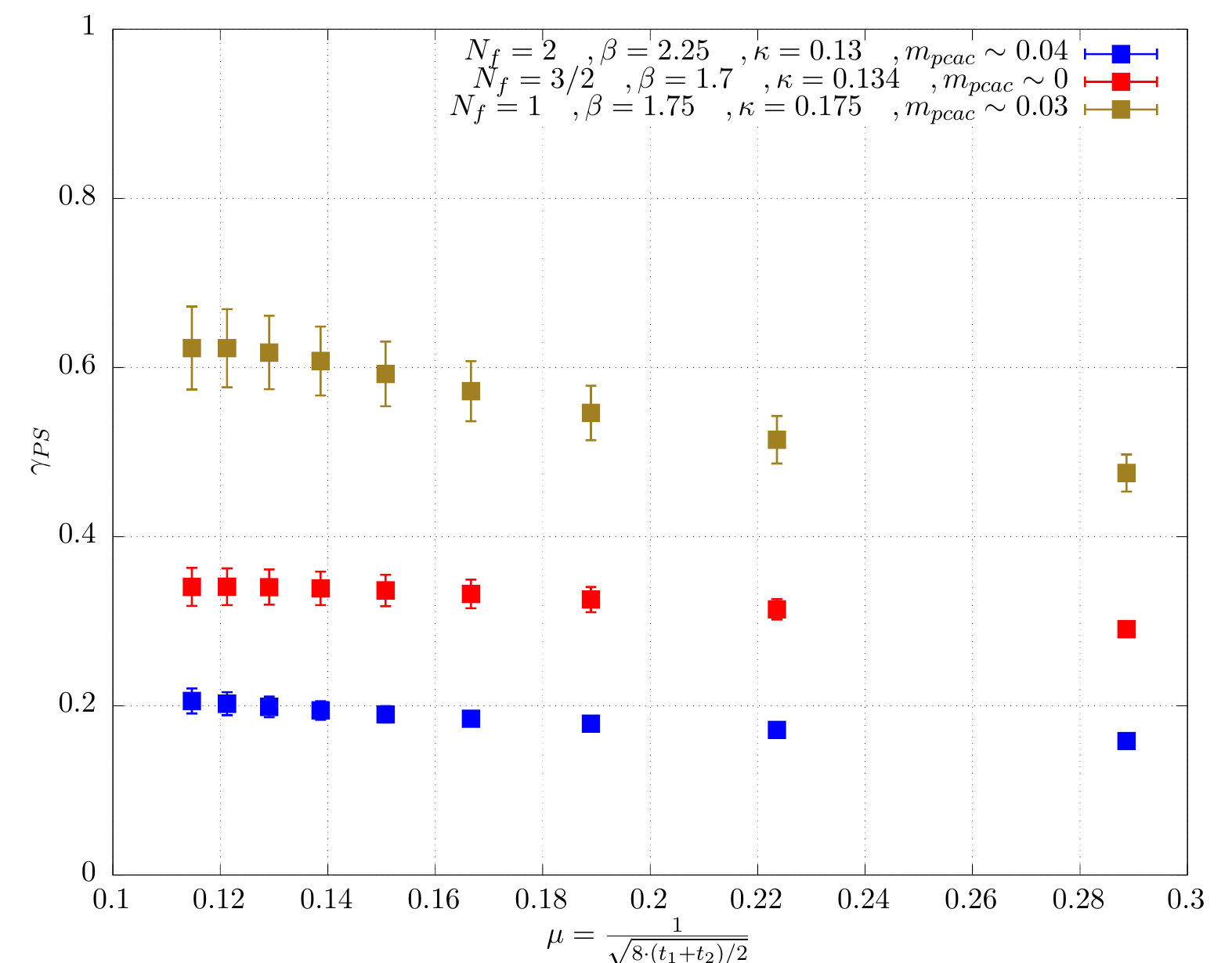}
\caption{\small Preliminary results of RG flow of $\gamma_{PS}$ for adjoint QCD with $N_{f}=1$, $3/2$ and $2$ Dirac flavors. The x-axis shows the energy scale $\mu=\frac{1}{\sqrt{8((t_{1}+t_{2})/2)}}$. The lattice volumes are $V_{4}=24^{3}\times 48$ for $N_{f}=1$ and $V_{4}=32^{3}\times 64$ in the other two cases.}
\label{nfcomp}
\end{figure}

The GF can be identified with the coarse-graining part of a RG transformation. Indeed, the GF kernel suppresses high momentum modes at finite flow time and smears fields over a radius proportional to the square root of the flow time. However, the GF is not a complete RG transformation since the high momentum modes are not integrated out and there is no rescaling of the lattice spacing like in a block spin transformation. Nevertheless, information of the RG flow can still be determined from the GF. To compute the anomalous scaling dimension $\gamma_{\mathcal{O}}$ of a given composite operator $\mathcal{O}[\phi]$, the RG transformation corresponding to a rescaling of the lattice spacing (UV cutoff) by a factor $b$ is considered. Since the GF is the coarse-graining part of the transformation, it is natural to assume $b\sim \sqrt{t}$.  The effect on a specific ratio of two-point functions is a scaling according to
\begin{align}\label{eq:ratio}
\frac{\langle\mathcal{O}_{t}(0)\mathcal{O}_{t}(x_{0})\rangle}{\langle\mathcal{O}(0)\mathcal{O}(x_{0})\rangle}=b^{2\Delta_{\mathcal{O}}-2n_{\mathcal{O}}\Delta_{\phi}}, \quad \Delta_{i}=d_{i}+\gamma_{i},
\end{align}
 where $\mathcal{O}_{t}$ is the flowed operator. The exponent is determined by the sum of  canonical ($d_{\mathcal{O}}$) and anomalous dimension ($\gamma_{\mathcal{O}}$). The operator  consists of products of $n_{\mathcal{O}}$ elementary fields, which provide the second term in the exponent. It is clear that $x_0$ has to be sufficiently large in order to be consistent with the coarse-graining.  We are interested in extracting $\gamma_{\mathcal{O}}$ from expression \eqref{eq:ratio}. To this end, we choose a conserved current $\mathcal{V}$, implying $\gamma_{\mathcal{V}}=0$, such that the scaling dimensions of $\phi$ are eliminated in the ratio
\begin{align*}
  \mathcal{R}_{\mathcal{O}}(t,x_{0})&=\frac{\langle\mathcal{O}(0)\mathcal{O}_{t}(x_{0})\rangle}{\langle\mathcal{O}(0)\mathcal{O}(x_{0})\rangle}\left(\frac{\langle\mathcal{V}(0)\mathcal{V}(x_{0})\rangle}{\langle\mathcal{V}(0)\mathcal{V}_{t}(x_{0})\rangle}  \right)^{n_{\mathcal{O}}/n_{\mathcal{V}}}=b^{\Delta_{\mathcal{O}}-(n_{\mathcal{O}}/n_{\mathcal{V}})d_{\mathcal{V}}}\propto t^{\gamma_{\mathcal{O}}/2+d_{\mathcal{O}}/2-(n_{\mathcal{O}}/n_{\mathcal{V}})d_{\mathcal{V}}/2}\, .
\end{align*} 
Choosing $\mathcal{V}$ to have the same field content as $\mathcal{O}$, i.e. $n_{\mathcal{V}}=n_{\mathcal{O}}$ and  $d_{\mathcal{V}}=d_{\mathcal{O}}$, we find the simple relation $\mathcal{R}_{\mathcal{O}}(t,x_{0})\propto t^{\gamma_{\mathcal{O}}/2}$. To determine $\gamma_m$ as a function of the scale given by $\bar{t}=(t_1+t_2)/2$, we compute the anomalous dimension of the pseudoscalar operator ($\gamma_{m}\equiv\gamma_{PS}$), 
\begin{align*}
\gamma_{PS}(\bar{t})=\frac{\log(\mathcal{R}_{PS}(t_{1})/\mathcal{R}_{PS}(t_{2}))}{\log{(\sqrt{t_{1}}/\sqrt{t_{2}})}},
\end{align*}
where $t_{1}$ and $t_{2}$ are two consecutive flow times. Since we consider Wilson fermions, the most natural choice for $\mathcal{V}$ is the conserved vector current ($\tau^a$ are the generators of flavor rotations)
\begin{align*}
\mathcal{V}^{a}_{\mu}(x)=\frac{1}{4}\left(\bar\lambda(x+\hat \mu)(1+\gamma_{\mu})U_{\mu}(x)^{\dagger}\tau^{a}\lambda(x)-\bar\lambda(x)(1-\gamma_{\mu})U_{\mu}(x)\tau^{a}\lambda(x+\hat \mu) \right)\, .
\end{align*}
We have computed $\gamma_{PS}$ for $N_{f}=1$, $3/2$, and $2$ flavors at different $\beta$,  mass parameters ($\kappa$) and volumes. We found that non-zero mass deformation become negligible around $m_{\mathrm{PCAC}}\sim 0.01$. Approximate scale-independence of $\gamma_{PS}$ is already seen at these small masses. Nevertheless, further effort is still needed to control $\beta$ and volume dependence to complete the extrapolation of the fixed point value $\gamma_{\star}$ in the deep infrared region. Our preliminary results already show qualitative hints for conformality in the case of $N_{f}= 3/2$ and $2$. The one flavor case might be interpreted as a nearly conformal behavior (see Figure \ref{nfcomp}). Indeed, for the two largest $N_f$ the value $\gamma_{PS}$ is nearly constant towards the IR, as expected for a fixed point. However, the precise values where the points seem to converge should be taken with care, as there is still a not yet determined $\beta$ dependence. In the specific case of $N_{f}=2$, the result in Figure \ref{nfcomp} at $\beta=2.25$ shows a rather small $\gamma_{PS}$ value with respect to other lattice studies (see for example \cite{Bergner:2016hip}). However, at $\beta=1.5$ we have observed values of the order of $\gamma_{PS}\sim 0.3$. We hope to resolve the remnant $\beta$ dependence in our following investigations.

\section{Phase structure of SU(3) super Yang-Mills theory}

The phase transitions of super Yang-Mills theory is particularly interesting since there are two independent order parameters related to chiral symmetry breaking and deconfinement. In addition  the symmetry constraints make it easier to study the QCD-like theory with analytical methods. The Lagrange density in the Euclidean space-time can be written as
\begin{align*}
      \mathcal{L}_{\mathrm{E}}=\frac{1}{4}F^{2} + \frac{1}{2}\bar\lambda(\slashed{D}+ m_{\tilde{g}})\lambda \,, 
\end{align*} 
where $F$ is the SU($N$) YM field strength and $\lambda$ is the gluino, a Majorana spinor in the adjoint representation. The mass is a soft SUSY breaking term, which is necessary for the lattice simulations. 
 Due to instanton effects, the $U(1)$ axial symmetry of the massless classical theory is is broken down to the discrete chiral subgroup $Z_{2N}$. 
 At lower temperatures, chiral $Z_{2N}$ symmetry is expected to be spontaneously broken down to $Z_{2}$ by a non-vanishing gluino condensate $<\bar\lambda\lambda>$. Domain walls appear as a consequence of the breaking of the discrete symmetry, interpolating between the $N$ degenerate vacua. It is known from analytical studies that the wall is BPS saturated and a Chern-Simons theory is expected to live on it. At larger temperatures the condensate is expected to melt leading to a restoration of chiral symmetry. 
 
 The theory is also invariant under global $Z_{N}$ center symmetry transformations. Like in Yang-Mills theory, this symmetry is spontaneously broken at the deconfinement transition with the Polyakov loop as the order parameter.
 
In our current studies we are considering the phase transition of SU(3) super Yang-Mills theory to extend our previous studies of the gauge group SU(2) \cite{Bergner:2019dim}. Contrary to QCD, center and chiral symmetries are exact in the massless limit and thus the Polyakov loop and the gluino condensate are valid order parameters. We have measured these observables on the lattice with clover-improved Wilson fermions (for further details, see \cite{Ali:2018dnd}). Non-zero temperatures are introduced by compactifying the Euclidean time dimension at fixed bare coupling $\beta$, imposing anti-periodic boundary conditions for the fermions. We have simulated lattices with $V=16^{3}\times N_t$ and $N_{t}$ between 4 and 12, for three different $\kappa$ values $0.165,0.1667$, and $0.1673$.

The Polyakov loop (PL) is computed in the standard way, while the definition of the gluino condensate is more subtle. The condensate requires additive and multiplicative renormalization when using Wilson fermions. Fortunately, the additive renormalization is eliminated for positive flow-times, as correlation functions of flowed local operators renormalize multiplicatively\footnote{The multiplicative renormalization can be fixed once a renormalization scheme is specified. It can be ignored in the determination of the phase transition.} \cite{Luscher:2013cpa}. On the lattice, the flowed gluino condensate is defined as
\begin{align*}
\langle\bar\lambda\lambda(t) \rangle=-\sum_{v,w}\langle{\mathrm{tr}\{K(t,x;0,v)(D_{\mathrm{W}}(v,w))^{-1}K(t,x;0,w)^{\dagger}\}\rangle}\,,
\end{align*}
where $K$ is the kernel of the fermion flow equation and $D_{W}$ the Dirac-Wilson operator. For details about the numerical computation, we refer the reader to reference \cite{Bergner:2019dim}.

\begin{figure}[h!]
  \centering
  \begin{subfigure}[l]{0.52\textwidth}
\includegraphics[scale=0.5]{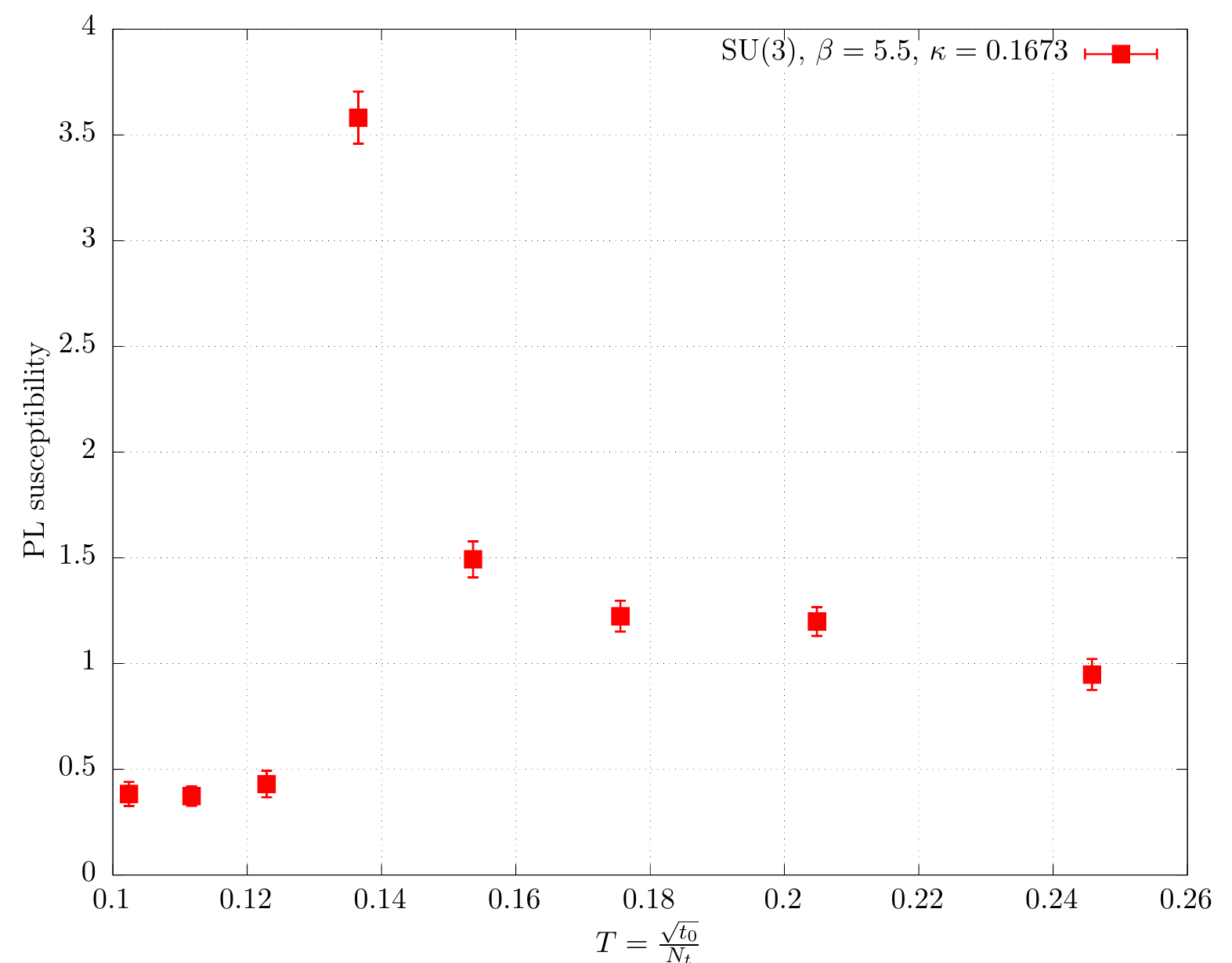}
\end{subfigure}
\begin{subfigure}[r]{0.47\textwidth}
  \includegraphics[scale=0.5]{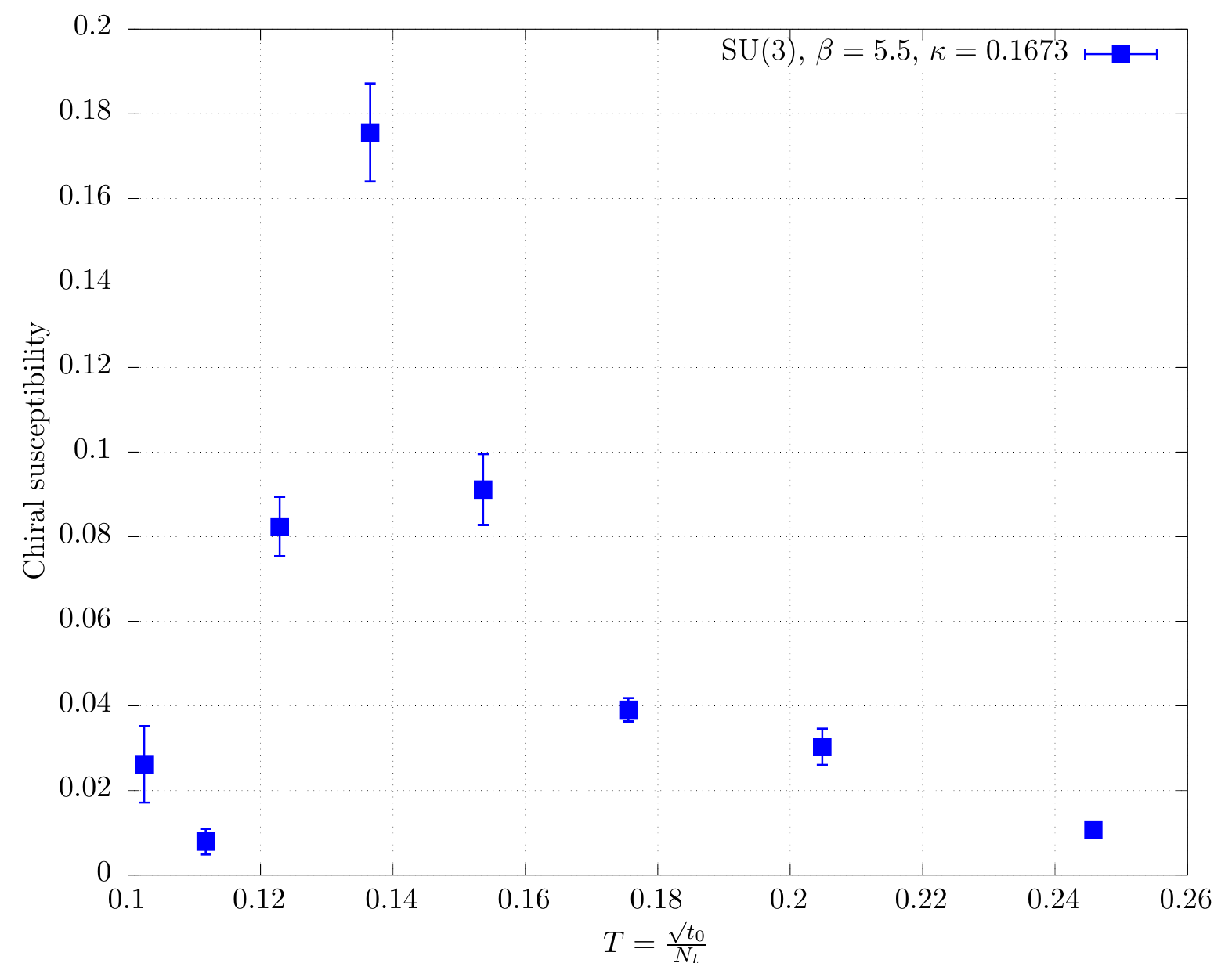}
\end{subfigure}
\caption{\small Polyakov loop and chiral susceptibilities (left and right resp.) for different temperatures. The critical temperature for both phase transitions seems to be $T_{c}\sim 0.14$, corresponding to $N_{t}=9$.}
\label{susc}
\end{figure}

In this proceeding we show the results for the smallest gluino mass corresponding to the largest $\kappa$, where the signal of the transition is stronger. In Figure \ref{susc} the chiral and Polyakov loop susceptibilities are presented. The temperature is in units of the GF scale $t_{0}$. These Figures suggest that both transitions coincide within the given uncertainty.

\begin{figure}[h!]
  \centering
  \begin{subfigure}[l]{0.52\textwidth}
\includegraphics[scale=0.5]{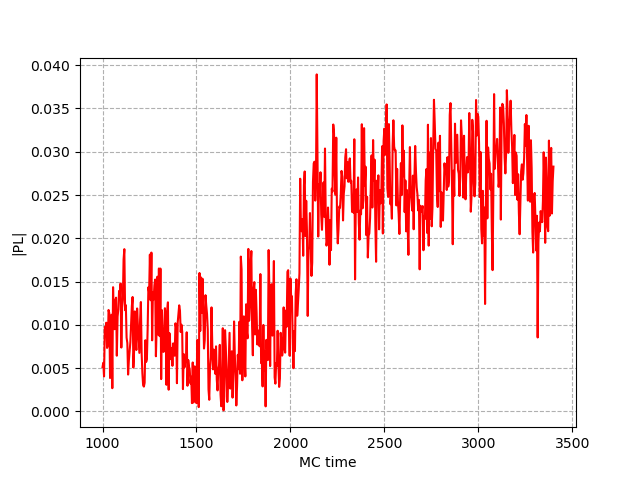}
\end{subfigure}
\begin{subfigure}[r]{0.47\textwidth}
  \includegraphics[scale=0.5]{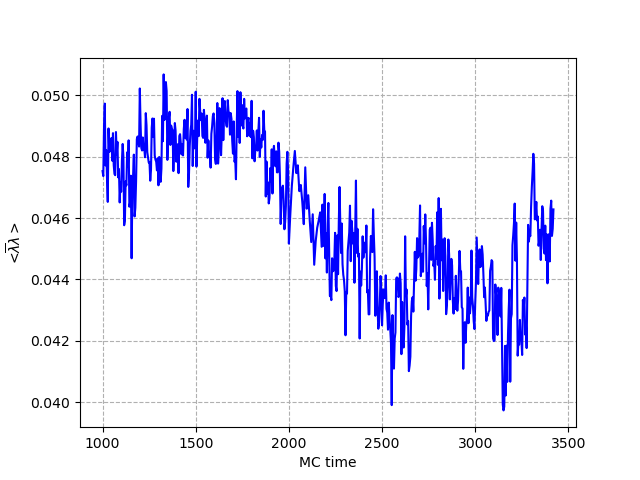}
\end{subfigure}
\caption{\small Monte Carlo histories of Polyakov loop (left) and gluino condensate (right) at $N_{t}=9$. }
\label{mch}
\end{figure}

Our present data is not sufficient to determine the order of the transition by finite size scaling. However, some indication for a first order transition can be deduced from the Monte Carlo histories close to the critical temperature ($N_{t}=9$), see Figure \ref{mch}. The Polyakov loop history suggests the coexistence of both center broken and center unbroken phases. A larger statistics is required in order to confirm this result, but it provides already evidence for a first order phase transition. For the gluino condensate a similar behavior is found. Moreover, we observe a correlation between the two order parameters. In other words, on a given configuration, a small Polyakov loop is correlated with a larger gluino condensate. 

\section{Conclusions and Outlook}

In this proceeding we have presented recent preliminary results regarding the phases of QCD(adj) with $N_f$ fermions in the adjoint representation. We have addressed two different questions using the methods based on the GF. 

First, we have exploited a relation between the GF and RG transformations as proposed in Reference \cite{Carosso:2018bmz} to compute the RG flow of the mass anomalous dimension for the theory with $N_{f}=1$, $3/2$, and $2$ Dirac flavors. Our preliminary results might be seen as a further evidence that the two largest flavor numbers are inside the conformal window. For $N_f=1$ a conformal behavior is not excluded. We are currently improving our results and expect soon to be able to determine a value for $\gamma_{\star}$ in each case. 
This quantity is important for considerations of Standard Model extensions. A further interesting investigation we would like to carry out in the near future is the computation of the continuous non-perturbative $\beta-$function as proposed recently in \cite{Hasenfratz:2019hpg}. This would clearly improve our knowledge on how adjoint QCD behaves at low energies.

Finally, we have used the gradient flow to study the thermal phase transitions in SU(3) SYM. Our preliminary results are in agreement with the intertwining of chiral and center symmetry. We observe indications for a single first order phase transition, where simultaneously the gluino condensate vanishes and the Polyakov loop gains a non-vanishing expectation value. These results are very similar to the SU(2) theory, where a single second order phase transition was observed.

\section{Acknowledgements}
The configurations of this study have been created by the DESY-Münster collaboration. We thank in particular Gernot M\"unster and Istvan Montvay for helpful comments and discussions.
The authors gratefully acknowledge the Gauss Centre for Supercomputing
e.~V.\ (www.gauss-centre.eu) for funding this project by providing
computing time on the GCS Supercomputers JUQUEEN, JURECA, and JUWELS at J\"ulich Supercomputing
Centre (JSC) and SuperMUC at Leibniz Supercomputing Centre (LRZ).
G.~Bergner and C.~L\'opez acknowledge support from the Deutsche
Forschungsgemeinschaft (DFG) Grant No.\ BE 5942/2-1.

\end{document}